# MODELLING RESILIENCE IN CLOUD-SCALE DATA CENTRES

**John Cartlidge** [a] **& Ilango Sriram**[b]

Department of Computer Science
University of Bristol
Bristol, UK, BS8 1UB

[a] john.cartlidge@bristol.ac.uk, [b] ilango@cs.bris.ac.uk

**ABSTRACT**
The trend for cloud computing has initiated a race towards data centres (DC) of an ever-increasing size. The largest DCs now contain many hundreds of thousands of virtual machine (VM) services. Given the finite lifespan of hardware, such large DCs are subject to frequent hardware failure events that can lead to disruption of service. To counter this, multiple redundant copies of task threads may be distributed around a DC to ensure that individual hardware failures do not cause entire jobs to fail. Here, we present results demonstrating the resilience of different job scheduling algorithms in a simulated DC with hardware failure. We use a simple model of jobs distributed across a hardware network to demonstrate the relationship between resilience and additional communication costs of different scheduling methods.

## 1. INTRODUCTION

Cloud computing—the online utility provision of hardware and software computing infrastructure and applications—necessitates the demand for data centres (DC) on an ever-increasing scale. The largest now fill purpose-built facilities approaching one million feet.[1] Already, DCs are so large that manufacturers (including IBM, HP, Sun) do not have the capability to build and destructively test models on the scale of the final production systems. Hence, every day, massively parallel, tightly-coupled, complex and sometimes heterogeneous data centres are put to service having undergone insufficient pre-testing; while it is still possible to test individual node servers and other standalone hardware, the complex interactions between the components of the DC under normal and abnormal operating conditions are largely unknown. Whereas in other engineering domains this problem has been addressed with robust industry-standard simulation tools—SPICE for integrated circuit design (Nagel 1975), or computational fluid dynamics for the aeronautics industry—a well established realistic (rigorous) simulation framework of cloud computing facilities is lacking.

There are two important reasons why this is the case. Firstly, there is no uniform definition of what a cloud computing infrastructure or platform should look like: where Amazon uses virtualization (DeCandia *et al.* 2007), Google uses MapReduce (Dean and Ghemawat 2008). Secondly, it is a hard problem: a realistic simulation tool should include real network models (fibre channel, Gbit ethernet), disk models (disk arrays, solid-state, caching, distributed protocols and file systems), queueing models for web servers, etc. As such, while it is our long-term goal to develop a set of simulation tools that can be used to aid the development of cloud DCs, as an initial step we present a tractable problem using a simplified model.

DCs for cloud computing have now reached such a vast scale that frequent hardware failures (both temporary and permanent) have become a normal expectation. For example, if a DC contains 100,000 servers and the average commodity server life expectancy is 3 years, we expect a server to reach the end of its natural life every 15 minutes; considering temporary failures and failure of other components makes failures occur even more frequently. Thus, when a job is submitted to the cloud, the physical hardware available at the start of the job cannot be guaranteed to be there at the end:

> *With such high component failure rates, an application running across thousands of machines may need to react to failure conditions on an hourly basis (Barroso and Hölzle 2009)*

To avoid frequent job failures, redundancy is necessary. The cloud computing design paradigm builds on achieving scalability by performing scale-out rather than scale-up operations, i.e., increasing resources by using additional components as opposed to using more powerful components. For this reason, jobs are generally split into parallel tasks that can be executed by (potentially) several services. For resilience purposes, the tasks can be multiply copied and run in parallel threads on different hardware (Hollnagel, Woods, and Levson 2006). Thus, as long as a "backup" copy exists, individual task failures will not degrade a job's overall resilience.

However, redundancy inherently generates extra work, requiring more space, greater computational effort and increased communication costs. There is clearly a trade off here: how much redundancy and how to schedule redundancy—where to physically locate copies of the same code in the DC to minimise the chances of failure—versus increased communication cost and computational effort.

---
[1] http://www.datacenterknowledge.com/special-report-the-worlds-largest-data-centers

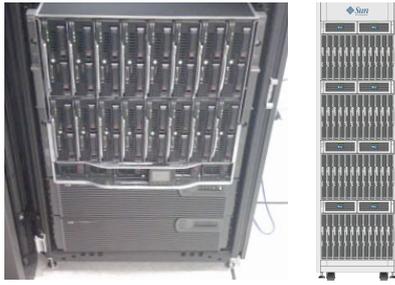

Figure 1: Example hardware: (a) HP C7000 chassis holding 16 blade servers; (b) Sun Pegasus C48 server rack, containing 4 chassis × 12 blade servers.

In this paper, we conduct an initial foray into the analysis of this trade off, using a simple simulation model to analyse the relationships between scheduling, redundancy, network structure and resilience. In Section 2 we introduce cloud-scale data centres and the problem of failure resilience. Section 3 outlines the simulation model we use, before detailing our experimental set-up in Section 4. Section 5 presents the results of our experiments, which are discussed in Section 6. In Section 7 we outline our future plans to extend this work, before summarising our conclusions in Section 8.

## 2. BACKGROUND

### 2.1. Cloud Data Centres

Cloud Computing transitions DCs from co-located computing facilities to large resources where components are highly connected and used in an interlinked way. Computations are broken down into services, allowing for easier scale-out operations. From the physical perspective, DCs are structured regularly in a hierarchical design: a warehouse scale DC is made up of aisles of racks, each rack being a vertical frame to which a number of chassis can be mounted; each chassis containing an arrangement of thin computer mother-board units: the blade-servers that make up the DC's computing infrastructure. Each blade server in turn hosts Virtual Machines (VMs) running cloud services. Figure 1 shows example chassis and rack components.

With Cloud Computing, the level of interconnectivity and dependency between services across the DC is so high that Barroso and Hölzle (2009) coined the term "warehouse-scale computers". This introduces various aspects of complexity to DCs. Firstly, many of the protocols in place scale worse than linearly, making conventional management techniques impractical beyond a certain scale as complex interactions between services lead to unpredictable behaviour. Secondly, DC design has reached a stage where test environments are no longer larger, or even of the same order of magnitude, as the final products. Cutting edge DCs are believed to have more than half a million cores,[2] but even one order of magnitude less would make a physical test environment too expensive to be practical. Hence, it is difficult to impossible to test the chosen configurations before going into production, which can lead to costly errors.

This highlights the need for predictive computer simulations to evaluate possible designs before they go into production: with simulation studies it is possible to rapidly evaluate design alternatives. However, for simulating cloud-scale computing DCs there are currently no well-established tools.

The literature includes some early-stage cloud simulation models. For a consumer centric view of the cloud, there is CloudSim (Buyya, Ranjan, and Calheiros 2009). CloudSim's design goal is to compare the performance of services on a cloud with limited resources against their performance on dedicated hardware. To aid the vendor perspective, we have previously developed SPECI (Simulation Program for Elastic Cloud Infrastructures) for modelling scaling properties of middleware policy distribution in virtualised cloud data centres (Sriram and Cliff 2011). This paper explores aspects of resilience modelling that we aim to develop as a component in a set of simulation tools for data centre designers.

### 2.2. Failure, Resilience and Communication Cost

As economies of scale drive the growth of DCs, there are such a large number of individual independent hardware components that the average life expectancy will imply that component failure will occur continually and not just in exceptional or unusual cases. This expected near permanent failing of components is called *normal failure*. For practicable maintenance, failed components are left *in situ* and only replaced from time to time; it may also be imagined that entire racks are replaced once several servers on it have failed. However, despite normal failure, resiliency must be maintained. Furthermore, the cloud design paradigm of solving jobs using smaller tasks or services that are typically spread across several physical components further increases the risk of normal failure affecting any given job. As cloud vendors seek to provide reliable services, requiring the maintenance of guaranteed levels of performance and dependability, resilience has become a new non-functional requirement (Liu, Deters, and Zhang 2010). To this end, cloud applications such as BigTable, Google's massively parallel data storage application, have in-built management systems for dealing with failures (Chang *et al*. 2008).

Hardware failure can occur anywhere in the physical hierarchy of the data centre: power outages can disable an entire DC; faulty cooling system behaviour can force an aisle to be shutdown to avoid overheating; racks, chassis and blades have individual power supplies which can fail; and individual VMs can suffer from instability in software and require an unplanned reboot. Thus, with growing DC scales, resources can no longer be treated as stable; and interactions no longer static but rather exhibiting dynamic interactions on multiple descriptive levels.

---

[2] http://www.zdnet.com/blog/storage/googles-650000-core-warehouse-size-computer/213

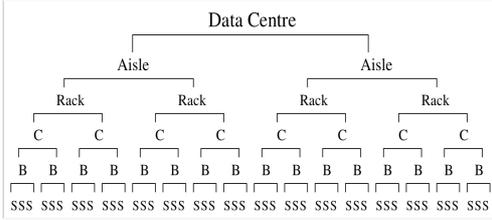

Figure 2: Data centre tree schematic. We describe this as an *h-2-2-2-3* hierarchy (2 racks per aisle, 2 chassis per rack, 2 blades per chassis and 3 services per blade). The full DC contains as many aisles as necessary.

To counter normal failure, redundancy must be introduced. This happens by spinning off parallel copies of all tasks. Thus, when any task from the original copy fails, a redundant copy is available to replace the service that has gone "missing". Hadoop, for example, is an open-source software for reliable, scalable distributed computing and is used by Yahoo!, Facebook and others, on clusters with several thousand nodes.[3] It includes HDFS file system, which as default creates 3 copies (redundancy 3).[4]

When considering parallel execution of tasks rather than file storage, service redundancy causes extra load through the additional execution of tasks. The execution load grows linearly with the numbers of redundant copies, but in addition, there will be some form of load associated with parallel threads periodically passing runtime data that we describe as communication cost. This paper uses a simulation model of parallel job execution to explore the trade-off between resilience and communication cost as failure, redundancy and scheduling types vary. For model simplicity we focus on computational redundancy and ignore disk and I/O redundancy.

## 3. SIMULATION MODEL

### 3.1. Network Tree Hierarchy
We model the interactions between networks of VM cloud services that exist in a hierarchical tree-structure (refer to Figure 2). Network structure is configurable and we use several tree hierarchies. Throughout this paper, however, unless otherwise stated assume a fixed hierarchy *h-8-4-16-16*. That is, each aisle has 8 racks, each with 4 chassis containing 16 blades, with each blade running 16 cloud services. This structure was chosen to model realistic hardware, such as the 16-blade HP C7000 chassis and 4-chassis IBM rack shown in Figure 1.

### 3.2. Jobs, Tasks and Redundancy
We assume that all jobs to be run in the cloud can be parallelized into $T$ independent task threads. We make this simplifying assumption on the basis that one of the major draws of cloud infrastructures is the elasticity of rapid scaling and de-scaling through parallelization. In our model, $J$ jobs are run on the DC, with each job, **J**, consisting of $T$ independent parallel tasks. While tasks can be parallelised, they are not entirely independent otherwise they would constitute a new job. Thus, tasks must periodically communicate with each other, passing runtime data when necessary. To pass runtime data, tasks within a job communicate at fixed time intervals. Normally, if any one task within a job fails, the entire job will fail. To mitigate this, redundancy can be introduced by running $R$ duplicate copies of tasks in parallel. Then, job **J** will fail *if and only if* all redundant copies of an independent parallel task fail. Such redundancy introduces failure resilience.

Let **J** denote a job consisting of $T$ tasks, each having $R$ redundant copies. Then, **J** can be written in matrix notation, with $T$ rows and $R$ columns:

$$J = \begin{pmatrix} j_{1,1} & j_{1,2} & \cdots & j_{1,r} & \cdots & j_{1,R} \\ j_{2,1} & j_{2,2} & \cdots & j_{2,r} & \cdots & j_{2,R} \\ \vdots & \vdots & \ddots & \vdots & \ddots & \vdots \\ j_{t,1} & j_{t,2} & \cdots & j_{t,r} & \cdots & j_{t,R} \\ \vdots & \vdots & \ddots & \vdots & \ddots & \vdots \\ j_{T,1} & j_{T,2} & \cdots & j_{T,r} & \cdots & j_{T,R} \end{pmatrix} \quad (1)$$

Job failure occurs when all tasks in a given row fail. More formally:

$$fails(J) \Leftrightarrow \exists t \in T, \{\forall r \in R, fails(j_{t,r})\} \quad (2)$$

Throughout this paper, we denote experiments running $J$ jobs, each with $T$ tasks and $R$ redundancy as a $\{J, T, R\}$ configuration, with sum total tasks $\#T = J \times T \times R$.

### 3.3. Scheduling Algorithms
Jobs and tasks can be placed onto a DC network in an infinite variety of ways; using schedules that range from the simple to the complex. In this work, we are interested in deriving general relationships between job scheduling methods and the effects they have on communication cost and resilience. Since we cannot hope to assess the relative behaviours of *every* scheduling algorithm, to aid analytical tractability, we selected a small subset purposely designed to be simple. The intention is not to test intelligent, complicated, real-world algorithms, but rather to tease out general behaviours of these simple algorithms so that we can equip ourselves with better knowledge to design intelligent industrial algorithms in the future. To this end, we define the following three scheduling algorithms:

---
[3] http://wiki.apache.org/hadoop/PoweredBy
[4] http://www.hadoop-blog.com/2010/11/how-to-change-replication-factor-of.html

- Random: Uniformly distribute tasks across the DC, independent of job or redundancy group.
- Pack: Use the minimum amount of DC hardware to run all jobs. Place tasks from consecutive redundancy groups for all jobs on consecutive DC services.
- Cluster: Place all tasks belonging to the same redundancy group on the smallest piece of hardware that they fit (e.g., on one blade).[5] Uniformly distribute redundancy groups across the DC.

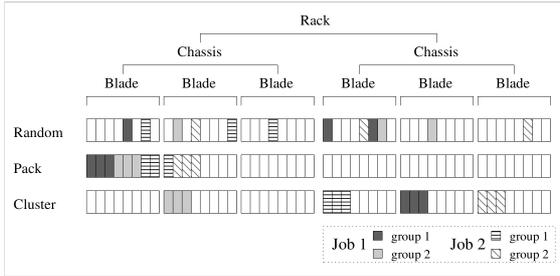

Figure 3: Job scheduling for a {*J*=2, *T*=3, *R*=2} configuration on an *h-2-3-8* hierarchy subset. Top: Random uniformly distributes the *#T*=12 tasks across the DC. Middle: Pack schedules tasks onto the minimum physical hardware set, in this case 2 blades on 1 chassis. Bottom: Cluster schedules full job copies onto the same physical hardware, while uniformly distributing copies across the DC.

Figure 3 shows a schematic example of each scheduling algorithm. Random, *top line of figure*, assigns tasks to the DC using a random uniform distribution over all DC services. Random schedules tasks independently, taking no account of which job or redundancy group a task belongs. Conversely, Pack preserves geographical co-location of individual tasks according to job and redundancy groupings, *middle*. Tasks are sequentially scheduled using neighbouring services until each hardware is filled. Finally, Cluster uses a combined approach, *bottom*. Similar to Pack, Cluster places all tasks belonging to a job redundancy group on the same physical hardware. However, redundancy groups themselves are uniformly distributed across the DC. In aggregate, these trivial scheduling algorithms form a simple strategy spanning-set from which we aim to tease out general rules for improving failure resilience.

### 3.4. Network Communication Costs

As explained in Section 3.2, the model assumes that tasks within a job need to communicate at fixed time intervals, passing runtime data between parallel threads. Table 1 shows inter-task communication costs within the DC. Intuitively, cost increases with physical distance between tasks, increasing in magnitude each time it is necessary to traverse a higher layer in the DC tree (Figure 2). From Table 1, tasks communicate with themselves and other tasks on the same service with zero cost. For tasks on different services on the same blade the communication cost is $C_S=10^0$; between different blades $C_B=10^1$; etc.

[5] In the case that no hardware has enough free space to fit the entire task-group, deploy as many tasks as possible on the hardware with the largest free space, then deploy the remaining tasks as "close" (lowest communication cost) as possible.

Table 1: Communication Costs

| Communication | Relative Cost |
|---|---|
| Inter-Service | $C_S = 10^0$ |
| Inter-Blade | $C_B = 10^1$ |
| Inter-Chassis | $C_C = 10^2$ |
| Inter-Rack | $C_R = 10^3$ |
| Inter-Aisle | $C_A = 10^4$ |

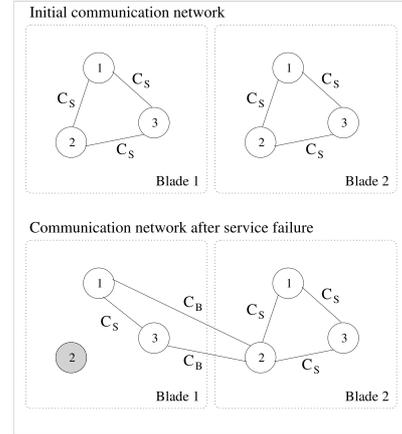

Figure 4: Communication network costs. Top: tasks communicate with the nearest copy of every other task. Bottom: when a task fails, communicating tasks find the nearest alternative. When task 2 fails, communication costs increase from $6C_S$ to $4C_S+2C_B$. Refer to Table 1 for cost values.

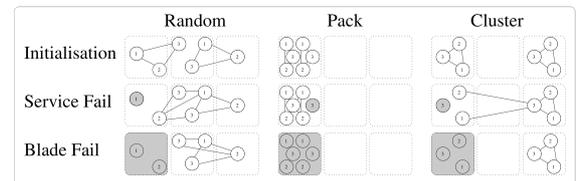

Figure 5: Hardware failure. Top: initial communication networks resulting from alternative scheduling methods. Middle: individual service failure produces minor restructuring of communication networks, including the addition of more costly inter-blade edges. Bottom: blade failure produces major network reconfiguration, while Random and Cluster recover Pack results in job failure.

These costs were chosen to give a qualitatively intuitive model of costs: clearly, communication costs between tasks running on the same physical chip are many orders of magnitude lower than between tasks located in different aisles of a data centre. While more accurate estimates of relative costs are possible, we believe the simple relationship defined in Table 1 adequately serves the purposes of this paper, since we are only interested in qualitative differences between scheduling algorithms, rather than accurate quantitative relationships.

### 3.5. Hardware Failure

Hardware failures directly affect task communication. Figure 4 highlights a schematic example of the effects of a single service failure. Initially, tasks communicate with the "nearest" copy of every other task; where "nearest" is defined as the least costly to communicate. Top: communication takes place between tasks running on the same blade. Middle: after task 2 fails on blade 1, tasks 1 and 3 on blade 1 begin inter-blade communication with the "nearest" alternative copy of task 2. The resulting network communication load increases from $6C_S$ to $4C_s+2C_B$.

Within the model, hardware failure can occur at any level in the physical hierarchy tree. Figure 5 demonstrates example effects of failure on the underlying communication network. Tasks form an initial communication network, *top*. Individual service failure, *middle row*, results in some restructuring of the communication network; with the addition of more costly inter-blade links for Random and Cluster. Hardware failure of an entire blade server, *bottom row*, has a more profound effect on the network. While Random and Cluster find a new rewiring, there no longer exists a full task-set for Pack, thus resulting in job failure.

### 4. EXPERIMENTAL DESIGN

Using the model described in Section 3, we perform a series of empirical experiments to observe the effect that different job scheduling algorithms have on resilience and communication cost in a data centre with hardware failures.

### 4.1. Assumptions

To keep the model tractable we make some simplifying assumptions.

*Time:* Simulations have a fixed time length. Jobs are scheduled before the simulation clock begins, then run for the entire length of the simulation.

*Jobs*: Jobs consist of a set of tasks that can be run in parallel with no inter-dependencies or I/O requests, other than periodic passing of runtime data. Consider, for example, computationally intensive batch jobs such as overnight computation of market data for financial institutions, or CFD simulations for the aeronautics industry or Met Office. For all tasks comprising a job, if at least one copy of the task succeeds, then the job completes successfully; refer to equation (2).

*Communication Cost*: Tasks within a job need to communicate with a copy of all other tasks at a constant rate. Communication costs increase with physical distance; see Table 1.

*Network Utilisation*: The DC is effectively infinite in size, enabling us to ignore the dynamics of full utilisation.

*Failure*: Failure can occur at any level in the hierarchical tree. Failure events are drawn from an exponential distribution.

### 4.2. Configuration

*Hierarchy Tree*: Unless otherwise stated, all experimental runs use an *h-8-4-16-16* tree hierarchy. These are realistic values based on current consumer hardware (refer to Section 3.1). Where alternative tree architectures are used, we use the notation *h-5* and *h-10* as shorthand for *h-5-5-5-5* and *h-10-10-10-10*, respectively.

*DC size*: To approximate unlimited resources, we scale the size of the data centre, |DC|, to equal twice the size needed to run all jobs, that is:

$$|DC| = 2 \times \#T = 2 \times J \times T \times R \qquad (3)$$

In an alternative configuration, data centre size is fixed. Under these conditions, set:

$$|DC| = 20 \times J \times T \qquad (4)$$

*Communication Costs*: Communication costs are set equal to Table 1 Refer to Section 3.4 for a discussion.

*Scheduling*: Jobs are scheduled using the algorithms Random, Pack and Cluster (as detailed in Section 3.3).

*Hardware Failure*: We set the proportion of hardware that will fail, $f_{hw}$, during the length of a simulation run to 1%, 5% or 10%. Note, however, that a failure event will cascade down the hierarchy tree, such that failure of a chassis will cause all blades and services running on the chassis to fail. Thus, the overall proportion of a DC that will fail during a simulation run will be larger than the value of $f_{hw}$. These failure rates may appear to be high. However, it is our intention to model resilience under extreme conditions that cannot be observed readily in operational environments. When a hardware failure event occurs, a discrete distribution is used to select the type of hardware failure. The relative probability of a given type of hardware, $h_{type}$, failing is calculated as the relative proportion of that hardware in the data centre, $h_{type}/h_{all}$. Although this distribution is simplistic, it provides the intuitive result that the more common a type of hardware, the more likely it is to fail.

### 5. RESULTS

Here, we present simulation results for all scheduling experiments. Figures plot mean values of repeated trials, plus or minus 95% confidence interval. Thus, where error bars do not overlap, differences are statistically significant. The simulation experiments were run in parallel and distributed across a cluster of 70 linux machines and the number of repetitions varies between 30 repetitions to over 100 repetitions. Confidence intervals remain relatively large due to the stochastic nature of the failure process: particular failure events can have widely ranging effects. It should be noted that occasionally the entire DC fails during simulation. When this occurs, the run is rejected so these catastrophic outliers do not skew results. This is reasonable since DC failure is a direct result of random hardware failure and is independent of the scheduling algorithms under test. Hence, all plots display summary data from trials where the entire DC did not fail.

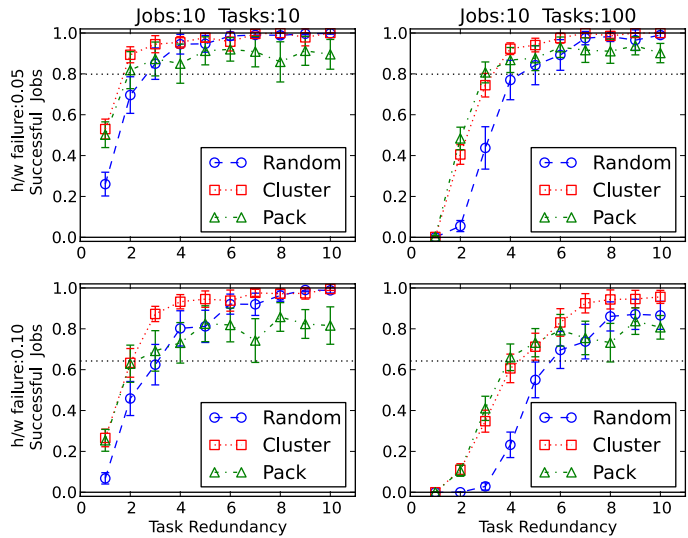

Figure 6: Resilience of scheduling algorithms in a fixed-size data centre with tree hierarchy *h-8-4-16-16*. As hardware failure increases, *top to bottom*, resilience falls; as tasks per job increases, *left to right*, resilience falls. Overall, Cluster is more resilient than Random and Pack across all conditions. Error bars show 95% confidence intervals. The dotted horizontal line plots the mean percentage of DC services surviving at the end of a run: the resilience that jobs with $T=1$ and $R=1$ will tend toward.

## 5.1. Resilience

Figure 6 shows simulation results for $J=10$ jobs using fixed DC size, equation (4). The proportion of successful job completions, $S_J$, is plotted against number of redundant task copies, $R$, for each algorithm: Random, Cluster, and Pack. In each graph, we see the intuitive result that success, $S_J$, increases with redundancy, $R$. However, whereas Random (blue circle) and Cluster (red square) reach 100% success under all conditions except bottom-right, Pack (green triangle) reaches a maximum in the range 80%-90% at approximately $R=5$ and then plateaus, with fluctuation, as $R$ is increased further. As shown schematically in Figure 5, Pack schedules all tasks to fit on the smallest hardware set possible. However, this tactic of "putting all your eggs in one basket" is vulnerable to specific hardware failure events that may take out the entire set of tasks. Although such events are rare, across all runs and all job sets they occur often enough to stop Pack from reaching $S_J$=100%, regardless of $R$.

For all algorithms, we see that as the number of tasks per job, $T$, is increased from $T=10$, *left*, to $T=100$, *right*, more redundancy is needed to maintain a given level of resilience. This is intuitive. Since task failure results in job failure, the greater the number of tasks per job, the greater the chances of any one job failing; hence, the greater the number of redundant copies needed to counter this failure. Similarly, when the probability of hardware failure, $f_{hw}$, is increased from 0.05, *top*, to 0.10, *bottom*, to maintain resilience redundancy $R$ must be increased. Once again, this is intuitive: as failure increases, so too does the likelihood of job non-completion.

Overall, across all conditions, Cluster is the most resilient. With low values of $R$, Cluster and Pack outperform Random. When $R \geq 7$, Cluster and Random outperform Pack. Further, there is no condition under which Cluster is significantly outperformed by either Random or Pack. Yet, there are several conditions under which Cluster significantly outperforms both alternatives. Thus, results suggest that Cluster is the most robust strategy. Interestingly, the default number of redundancies used in Hadoop's HDFS, $R=3$, appears to be a reasonable choice when $T=10$. As the number of tasks increases, $R=3$ does not suffice under our conditions.

## 5.2. DC Architecture

Figure 6 displays a clear relationship between increased redundancy, $R$, and increased resilience, $S_J$. However, the resilience graphs for Pack exhibit "dips", for example at $R=8$ (*top-left* and *bottom-right*) and $R=7$

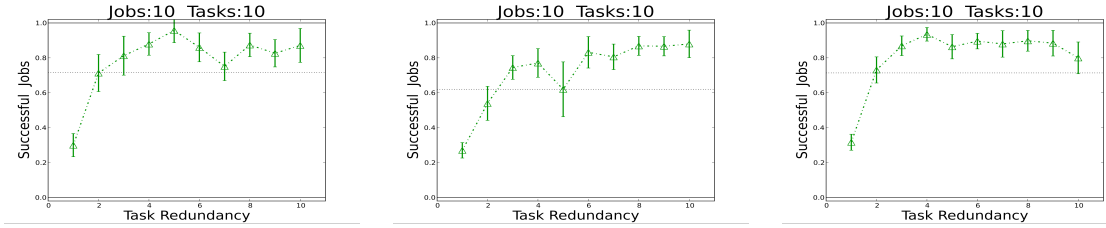

Figure 7: Pack scheduling using *h-8-4-16-16*, *h-5* and *h-10*, from left to right, respectively. Resilience dips when jobs exactly fit on underlying hardware.

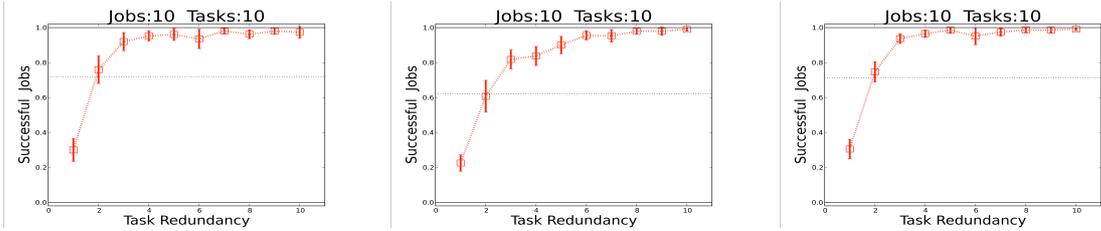

Figure 8: Cluster scheduling using *h-8-4-16-16*, *h-5* and *h-10*, from left to right, respectively. Cluster is largely insensitive to underlying hardware architecture.

(*bottom-left*), which raises some questions. Are these "dips" merely statistical aberrations, or are there underlying dynamics causing specific $R$ values to result in lower $S_J$?

To address this issue, a series of experiments were performed using alternative DC tree structures (Figure 2) to observe the effect that hardware hierarchy has on resilience. Three configurations were tested: *h-8-4-16-16*, *h-5* and *h-10*. In each case, data centre size |DC| was variable; refer to equation (3).

For all hierarchy trees, results were qualitatively similar to those displayed in Figure 6, suggesting that the general behaviours of each scheduling algorithm are largely insensitive to the underlying hardware hierarchy configuration. However, Pack does display some idiosyncratic sensitivity to hierarchy. Figure 7 plots $S_J$ against $R$ for Pack under each tree hierarchy. Left: $S_J$ increases with $R$ until $R$=5, but then fluctuates around 80%, with a minimum at $R$=7. When DC hierarchy is changed to *h-5*, *centre*, Pack has a minimum at $R$=5. Finally, with hierarchy *h-10*, *right*, there is a minimum at $R$=10. This evidence suggests that Pack is sensitive to the underlying physical hierarchy of the DC. In particular, if all redundant copies of a job fit exactly onto one hardware unit—blade, chassis, rack, etc.—then a failure on that hardware will take out all copies of the entire job. Hence, with an *h-5* hierarchy, for instance, $R$=5 results in poor resilience for Pack. Under these circumstances, each job, $J_{T,R}$, contains 50 tasks, which exactly fit onto 2 chassis. Thus, two neighbouring chassis failure events will take out the entire job. In comparison, when $R$=4 or $R$=6, task group copies will be more unevenly distributed over hardware, giving greater resilience to failure.

Figure 8 plots results for Cluster under the same conditions as Figure 7. Here, we see that Cluster is largely unaffected by the underlying tree structure of the data centre.

### 5.3. Jobs
To see how results scale with an increase in jobs, we ran experiments with $J$=100 jobs, using hierarchy *h-8-4-16-16* with variable data centre size, |DC|. Results are qualitatively similar to those of Figure 6. Pack outperforms Random with low $R$, but plateaus around $S_J$=80%. Random has poor resilience when $R$ is low, but outperforms Pack as $R$ approaches 10. Finally, Cluster has best resilience overall. Results for other failure rates and number of tasks, $T$, are also qualitatively similar, indicating that resilience is insensitive to $J$.

### 5.4. Communication Costs
While it is important for scheduling algorithms to enable job completion, it is also important they do not induce prohibitive cost through wasteful communication. In this section, we explore the mean communication cost, $C_J$, for each successfully completing job. Results suggest that the relationships between algorithms are largely robust to variations in hierarchy-tree, number of jobs, $J$, number of tasks, $T$, and hardware failure rate, $f_{hw}$. Thus, in this section we consider only two conditions: variable |DC|; and fixed |DC|. Each time an *h-8-4-16-16* architecture is used.

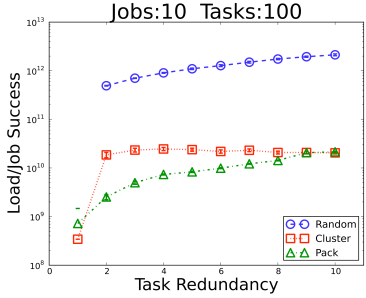

Figure 9: Costs per successful job, $C_J$, using fixed |DC| with $h$=8-4-16-16.

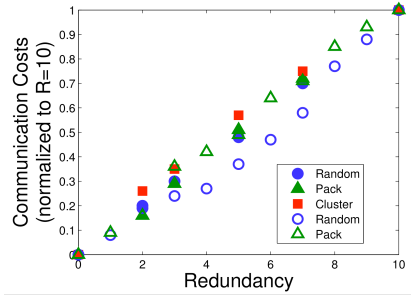

Figure 10: Communication costs per successful job, normalized to $R$=10. Clear-faced markers represent variable-sized DC; filled markers show fixed-size DC. We see that communication costs scale linearly with redundancy for all algorithms.

Figure 9 displays communication costs per successful job, $C_J$, using fixed |DC|. 95% confidence intervals are drawn, but are generally very small. With tasks uniformly distributed throughout the DC, Random produces the greatest communication costs per job. Conversely, with tasks placed as close together as possible, Pack has the smallest communication costs per job. For Pack and Random, an increase in redundancy, $R$, leads to a proportional increase in cost, $C_J$. When $R$=10, $C_J$ is approximately 10 times greater than the value at $R$=1.

For Cluster, however, the story is different. When $R$=1, Cluster produces smaller $C_J$ than Pack, since Cluster guarantees all job copies are placed on the lowest branch of the hardware tree that they fit; with Pack, however, depending upon number of tasks, $T$, and the underlying tree-hierarchy, some jobs will occasionally be split across hardware (see schematic Figure 3, for example), thus incurring greater communication costs. When redundancy is increased to $R$=2, communication costs, $C_J$, becomes a magnitude greater than Pack. As Cluster distributes job groups across the network, when an individual task fails, new communication links to alternative copies are likely to be long-range and costly. In contrast, Pack places all clones near each other, so communication with alternatives does not radically increase costs in most cases (refer to Figure 5). Interestingly, with fixed |DC| mean communication cost per successful job, $C_J$, remains constant when $R \geq 2$. Since Cluster distributes job copies uniformly across the data centre, the mean distance or cost for communication between tasks in different redundancy clusters is inversely proportional to $R$. Hence, additional redundancy reduces the mean communication cost a task must pay to communicate with an alternative clone, thus making $C_J$ invariant under changes to $R$. It should be noted that the same is not true for Random, however. Unlike Cluster, since Random distributes all tasks uniformly independent of redundancy group, the majority of communication paths are inter-hardware and costly. Hence, doubling $R$ will approximately double $C_J$.

When using a variable-sized data centre—equation (4)—results for $C_J$ against $R$ are similar to Figure 9 for Random and Pack. For Cluster, however, $C_J$ is no longer invariant to $R$ and instead increases proportionally as $R$ increases. As |DC| increases with each increase in $R$, the mean length between communicating tasks remains stable. Thus, as the number of tasks increases so too does overall communication costs.

Figure 10 plots normalized communication cost for each scheduling algorithm against redundancy, $R$. Clear faces show fixed |DC| (not including Cluster) re-plotted from Figure 9. Coloured faces show data from the equivalent set of runs using variable |DC|. In all cases, with all algorithms, there is clearly a linear relationship, suggesting that communication costs rise in direct proportion to $R$. Note, however that this is not the case for Cluster under fixed |DC| (not plotted): here, communication costs are invariant in $R$.

### 5.5. Summary of Findings
The main findings can be summarized as follows:

1. The network hierarchy tree has little effect on the resilience of scheduling algorithms (except in the case of Pack, where particular tree configurations have negative impact on particular levels of redundancy).
2. Cluster is the most resilient scheduling algorithm from the selection modelled. In contrast, Pack is a non-resilient high-risk algorithm.
3. Pack is the most efficient algorithm, Random the least. Cluster generates intermediate costs, but scales well under fixed data centre size.
4. Overall, Cluster is the most practical algorithm, effectively combining the efficiencies of Pack with the resilience of Random.

### 6. DISCUSSION
The aim of this work is to build an understanding of the general relationships between scheduling, resilience and costs (rather than perform a detailed analysis of any particular algorithm), the results presented support our basic endeavour to use simulation models as a methodological framework to design and test tools for elastic cloud-computing infrastructures. We do not suggest that Random, Pack or Cluster are practical job-schedulers that should (or would) be used in industry, but rather that these purposely naïve algorithms provide a simple base-line set of strategies that enable us to tease out fundamental relationships between density, clustering and spread of jobs; and the impact each has on resilience and communication cost. By using simulation to better understand how these concepts interact, we gain access to a powerful off-line test-bed for algorithm development that provides a design route towards more robust and efficient cloud-computing services. The simulation model we have used makes some simplifying assumptions that should ideally be relaxed. However, despite this, the model is powerful enough to highlight how underlying complex interactions, such as between scheduling and the shape of the hierarchy-tree, can affect resilience. This is a promising indication of the value of pursuing the goal of creating an extensible simulation framework for cloud computing.

## 7. FUTURE WORK

Here, we outline potential future extensions:

1. More realistic modelling assumptions: the introduction of sequential task inter-dependencies, heterogeneous jobs and services, full-DC utilisation, etc.
2. Model verification and validation using real-world data. Retroactive validation of results through testing on real-world DCs.
3. Introduction of other scheduling algorithms from industry and the development of novel algorithms using evolutionary computation as an automated design tool.
4. Monitor individual failure events rather than failure over time, to observe how the system changes when failure occurs, and what exactly takes place at this lower level of description.
5. Compare the effects of scale-up versus scale-out: If the resource usage increases, what does it mean for the resilience if more services are used, rather than more powerful ones?
6. Introduce migration of services to the scheduling algorithm. This allows a task to be cloned when a parallel instantiation fails, and the clone can then be migrated towards the other tasks belonging to that job.

## 8. CONCLUSIONS

We have presented a simulation model for testing the effects that different scheduling algorithms have on the resilience and communication costs of jobs running "in the cloud" of a large scale data centre (DC). Modelling the data centre as a tree-hierarchy of physical machines and jobs as a collection of parallel tasks that can be cloned, we have demonstrated the effects that different job-scheduling algorithms have on network resilience and communication cost. As intuition would expect, Packing all tasks together in a small area of the DC greatly reduces communication cost but increases risk of failure. Conversely, a Random distribution of tasks throughout the DC leads to greater resilience, but with a much elevated cost. Clustering tasks together in cohesive job-groups that are then distributed throughout the DC, however, results in a beneficial trade-off that assures resilience without prohibitive costs. This work provides a teasing glimpse into the powerful insights a cloud simulator can provide. Given the grand scale of the challenge, this work has naturally raised many open questions and introduced scope for future extensions.

## ACKNOWLEDGMENTS

Financial support for this work came from the EPSRC grant:[6] EP/H042644/17 (for J. Cartlidge) and from Hewlett-Packard's Automated Infrastructure Lab, HP Labs Bristol (for I. Sriram). The authors would like to thank Prof. Dave Cliff and the sponsors for their support and interest in this topic.

## AUTHORS BIOGRAPHY

Dr **John Cartlidge** is a Research Associate in cloud computing. His research interests include simulation modelling, automated trading and electronic markets, and evolutionary computing. **Ilango Sriram** is a final year PhD student soon to defend his thesis on simulation modelling for cloud-scale data centres.

---

[6] http://gow.epsrc.ac.uk/ViewGrant.aspx?GrantRef=EP/H042644/1